%% file: corruptionModel.tex
\documentclass[runningheads,a4paper]{llncs} 
\usepackage{epsfig}
\usepackage{multirow}

\usepackage{amsmath}
\usepackage{amssymb}
\usepackage{graphicx}
\usepackage{times}

\newcommand{\Aniti}{\mathcal{ANITI}}

\newcommand{\fsf}[1]{{\small\textsf{#1}}}

\begin{document}

\title{Tackling Corruption With Agents \& ICT: A Vision}

\titlerunning{Tackling Corruption With Agents \& ICT: A Vision}

\author{Biplav Srivastava}
\institute{IBM Research}



\maketitle




%
%

%


%
%



\begin{abstract}
Corruption is universally considered as an undesirable characteristic of public services.
But despite attempts to simplify and automate such services, there is very little prior work
to detect, handle or prevent corruption in computation literature. This is particularly surprising
given the significant advances made in detecting fraud, a relevant but different form of
public impropriety afflicting businesses. In this paper, 
we paint a vision of how information and communications technology (ICT),
and specifically agent-based methods, can help tackle corruption in public services and 
identify challenges to achieve the vision.

\end{abstract}






\input{introduction}
\input{motivation}

\input{vision}

\input{aniti}

\input{challenges}

\input{solution}
\input{conclusion}

\section{Acknowledgements}

The author will like to thank Nidhi Rajshree and Nirmit Desai for discussions 
on various aspects of the subject that has helped evolve this work.


\bibliographystyle{splncs03}
\bibliography{corruption}

\end{document}

%% file: introduction.tex
\section{Introduction}

Corruption afflicts both public
\cite{corruption-review,corruption-credit-formalmodel} and private
\cite{walmart} services world wide. 
 In this paper, we follow
the definition by the World Bank and the United Nations Development
Program (UNDP) \cite{un-corruption}, which have defined corruption as
the {\em "misuse or the abuse of public office for private gain"}
\cite{UNDP-corruption08}.  The manifestation of misuse can occur in
many ways, such as bribery, extortions, kickbacks, vendetta, and more.
The notion of corruption bears many facets, from ethical to economical
\cite{corruption-intro,corruption-review}. It is known that it has a
significant negative impact on the growth of economies and hence, is
universally considered undesirable\cite{why-corruption}. A popular definition of corruption is 
{\em Corruption = Monopoly + Discretion - Accountability}\cite{controlling-corruption}.

However, little is known about corruption in a computational sense making 
even detecting it hard let alone preventing or addressing it. Consider the questions:

\begin{itemize}
\item exchange of money: can a service for which the customer does not pay a fee
     (free service) be termed corrupt? Or conversely, can a 
      corrupt practice only happen if the customer pays for a service?
\item human agents: can a service be corrupt if the agent delivering
      the service is not a human but an automated agent? 
\item contention for resources: can corruption happen if delivering it
  requires no contention of resources? Alternatively, if resources are
  scarce, will an objective way of allocating them help remove
  corruption?
\item unique request: can a service be corrupt if it is one-of-a-kind and hence
         only requested once?
\item familiarity between service requester and provider: if a service's requester and provider 
         are known, does it definitely promote corruption? Conversely, if service requester or provider
         is anonymized, will this promote or reduce corruption? If multiple service requesters are 
         known to each other or different actors while delivering services,
         can this help reduce corruption?
\item accountability: if a service request was found to be handled in a corrupt manner, 
        can the culpability of agents involved or service requesters  be established legally?
        What traceability mechanisms reduce and what can promote corruption? 
\item brokers: does allowing brokers to mediate service requests on behalf of service requesters
         promote corruption?
\end{itemize}

Specifically, what are the ingredients necessary for a service to be
corrupt?  If these were known, they can be checked for easy corruption
detection and even prevention.  What are the features in a service process design that promote
corruption? If these were known, services could be better re-designed
to address them.  What are the agent behaviors (requester or service provider) 
that  incentivize corruption? If these were known, the agents could
be made more aware. What are the patterns of corruption and 
prescriptions on how they may be addressed?  If one considers popular
attempts to address corruption, automation and transparency via open
data are frequently mentioned\cite{corruption-opendata}. Why do 
they help and what more could be done? 

The goal of this paper is draw attention of the AI and Agent community to the
opportunities in making significant breakthroughs to tackle corruption by
using and enhancing their techniques. We next present an illustrative example 
and then paint a vision of less corruptible public processes. We then identify challenges
on the path to achieve it using agent technologies and conclude with a discussion.

A term often used for imprompriety in business  is {\em fraud} which is different from 
corruption. Phua et al give a comprehensive survey of fraud detection techniques
in \cite{fraud-survey} by considering  fraud as {\em the abuse of a profit organization's
system without necessarily leading to direct consequences}. The authors provide a 
heirarchy of fraud perpetrators from internal and external perspective of a firm
and describe various data mining approaches in literature to detect them.
Note that fraud has a direct monetary aspect which corruption does not and further, the victim 
of fraud may be an organization while that of corruption is usually the service requester. Consequently,
fraud is reported and investigated by organizations who spend extensive resources to tackle
them while corruption is reported by
individual service requesters to which organizations respond to.

%% file: motivation.tex
\section{Motivating Example}

\begin{figure*}
    \centering
    \includegraphics[width=1\columnwidth]{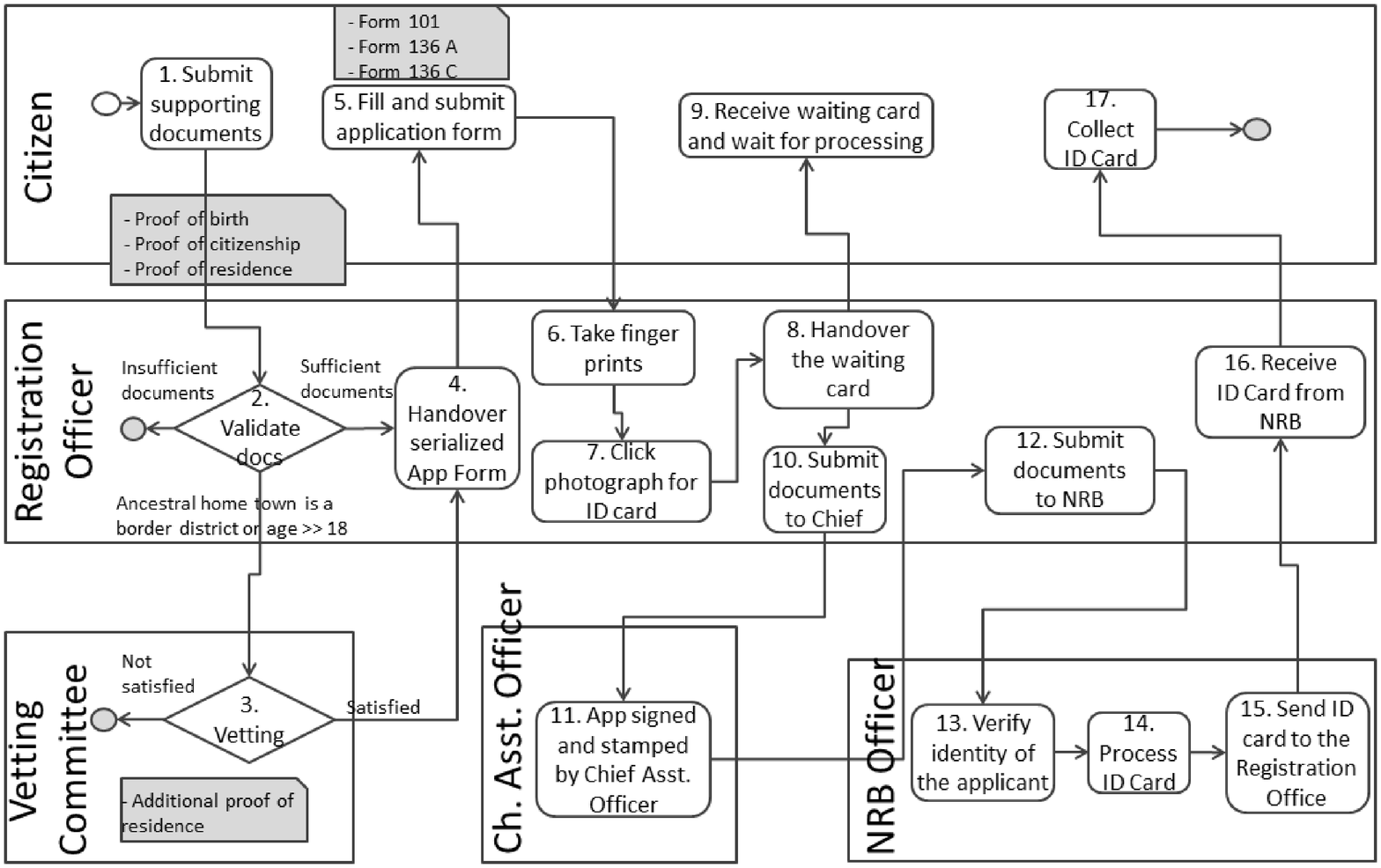}
    \caption{Identity Verification process in Kenya}
    \label{fig:id}
\end{figure*}

Our motivating example is the national registration and indentity verification
public process prevalent in almost all countries. This process leads to issuance of
a document (e.g., identity card, passport) without which, a person cannot request 
some or all of the following common trasactions: purchase or
own property, own a mobile phone, open a bank account, access higher
education, get employment, get subsidy entitlements or exercise the right to vote.
We illustrate with select examples of particular registrations in three countries -
Kenya (national identification), USA (social security) and India (Aadhar card)\footnote{Note that
a country may have multiple registration schemes covering eligibility of different services. So
the examples for the 3 countries are not exclusive.}.

%

The National identification (ID) card is the main form of identity
document in Kenya for anyone above the age of 18. It is also the
primary document to prove citizenship and is used to access many
public services. 
The process of registration for the National ID card begins by
applying at a division office anywhere in the country with a proof of
birth, citizenship and residence. In each category, there is a list of
acceptable documents. Interestingly, each category allows a signed
attestation by a high-level official to be acceptable as well. Once
the documents are found in order, further forms are submitted by the
applicant. During the process, finger prints are taken and the
documents are validated. Thereafter, an identity card is created and
sent to the applicant. 

The process is characterized by extremely wide discretion given to
registration officers in determining validity of documents. 
Specifically, the decision node, {\em 3 - vetting}, and the activity,
{\em 13 - verify identity}, are discretionary with no clear mechanism on how to accomplish them
but in contrast, the checks for documents having been submitted are objective.
There is no specification for service cost and time, and there is no escalation mechanism.
This, combined with the importance of ID card, leaves room for abuse,
arbitrariness and unreasonable demands on the applicants. 

Now consider the national registration processes in USA, called social security (SS) \cite{ssn-us},
which entitles and identifies citizens and non-citizens while taking government services. Here, 
a clear list of documents proving
US citizenship or legal residence for non-citizens, age and identity is listed. 
There is little room for discretion
because no category allows a signed attestation by a high-level official to be acceptable - something
allowed in Kenya. The cost and time limits for the service are prescribed. The process, however, 
can only be handled by a single agency creating a monopoly.

Now conside the case of  Aadhar card in \cite{aadhar-india} introduced recently to uniquely
identify a person using bio-metrics like finger prints for the purpose of delivering government's 
social services. 
To apply for this card, 18 Proofs of Identity (PoI) and   
33 Proofs of Address (PoA) documents are permitted. The process also allows discretion
by allowing attested documents from high-level officials. The cost and time limits for the service are prescribed. The process, however, can only be handled by a single agency creating a
monopoly. The federal government seeks to make all public services accessible contingent on a person
having an Aadhar card. Such a change can increase the card's importance but combined 
with wide official discretion, can have consequences on corruption.
This controversial move is being legally challenged in courts.

Countries and citizens of these and other nations may be interested to know which 
national identification processes are 
prone to corruption, to what degree and why. They may also want to take corrective
actions at design stage by re-designing the processes, at implementation stage by accounting for 
the issues in computation-based service delivery workflows,  at operations stage by defining redressal mechanisms or at periodic review stage. They would also want to learn from each other's experiences
and incorporate to improve their processes.

Apart from this process class, the corruption issue generalizes to all public processes. Hence
a significant challenge.

%
%


%% file: vision.tex

\section{Vision for Achieving Less Corruptible Public Processes}
\label{sec:vision}

We now paint a vision of how public processes may be made less corruptible in future
using information and computing technologies (ICT) in general and agent techniques
in particular. We see this as an evolution of progress.

\noindent {\bf Stage 1: Identify corruptible behavior}

The starting point for tackling corruption computationally is to have the ability to detect 
it unambiguously. The UNDP definition of corruption (i.e., {\em abuse or misuse ...})
is not helpful since it is hard to detect whether an agent is making private gain while
delivering a service. 

The known mechanisms for detecting corruption are non- computational - a) survey based, where public 
opinion surveys is used to measure the perceived corruption or b) institutional diagnostics
 such as audits and 
anti-corruption policy conformance of a public institution\cite{corr-tools}. 
The drawback of survey is that they are subjective and prone to human bias while
institutional diagnostics only assess the administrative loop holes and do not take care of scenarios where beneficiaries collude in corrupt practices.

A more pragmatic and computational approach was first suggested in \cite{corruption-semcity13}) 
and will be discussed in next section where 
a process is deemed suspicious (possibly corrupt) if given two same or similar service requests (inputs), 
they lead to different outcomes and hence outputs (functionally or non-functionally).
The assumption is that corruption-free processes are unbiased, and in unbiased processes, 
similar inputs should lead to the similar outcomes in terms of their
results (functional outputs) and also the time and resources incurred to get those 
results (non-fuctional outputs).

If one reaches stage 1 and can detect corruption, one will also like to be able to 
compare it across different 
processes. To illustrate, in the running example with
national processes of three different countries, which instance is more corruptible, 
Social Security in US or National ID in Kenya? Further, what specifically is the reason
for the process being corruptible ?

\noindent {\bf Stage 2: Build corruption patterns}

There is a large body of work on corruption in social sciences\cite{controlling-corruption},
\cite{nrega-visibility}.
From this, 
some root causes of corruption have been 
documented \cite{corruption-semcity13} .
A few are: actors being in a position of monopoly, service providers not having service level 
agreements (SLAs) within which to respond, actors having discretion to take decisions
without giving reasons, processes not having mechanisms to review an actor's decisions and
the requester not having visibility to their request status or provider's updates.

One would want to use the corruption detection models of stage 1 and build patterns
 of corruption root causes through which an agent can unambiguously
detect corruptible service request instances and their likely root causes. Over time, one
can also record what methods actually worked to tackle which root causes so that suitable
decision-support can be built (see next stage).

\noindent {\bf Stage 3: Recommend corruption reducing measures}

In terms of tackling corruption, there is  a belief that if
processes were automated or more data about service
delivery would be made available\cite{corruption-opendata} or processes were re-designed, 
this would help
help tackle corruption in public services. But why do these approaches seem to
work?  For example, as observed in \cite{autoseek,corruption-semcity13}, automation
requires formalization of input data requirements and specification of outputs. It is
possible that the precise specification of inputs and outputs, which removes discretion,
is the actual reason for perceived reduction in corruption due to automation. Similary,
open data brings more information about service outcomes into public domain
and thus allows faster detection of corruption. Process re-design may help reduce
corruption by removing unnecessary complexities and promoting traceability. There may be
more methods to control corruption and reasons why they work.

So, if there is a corruption cause seen in a process, what measure should one use for
that particular process type? Further, what is the least 
change needed in the corruption-prone processes or service delivering organizations to 
prevent corruption? 

We believe that there is an opportunity to build decision support to help make
processes less corruptible.

\noindent {\bf Stage 4: Prevent corruptible processes}

The eventual success in tackling corruption will be when corruptible processes are 
prevented from  getting executed. To do so, corruptible behavior should be detected
when the process is getting designed and flagged. If they are still pursued, measures should 
get introduced in their implementation and operation that initiates actions (e.g., automatic reporting 
of suspect behavior) when requests go against uncorrupt baseline.

Eliminating corruption will remain hard as long as detecting private gains of actors goes undetected.
However, to the extent observable behavior of  public processes are formalized, suspicious
behaviors can be detected and tackled.


%% file: aniti.tex
\section{A Formal Vocabulary for Corruption and Its Detection}
\label{sec:aniti}

\begin{figure*}
    \centering
    \includegraphics[width=1\columnwidth]{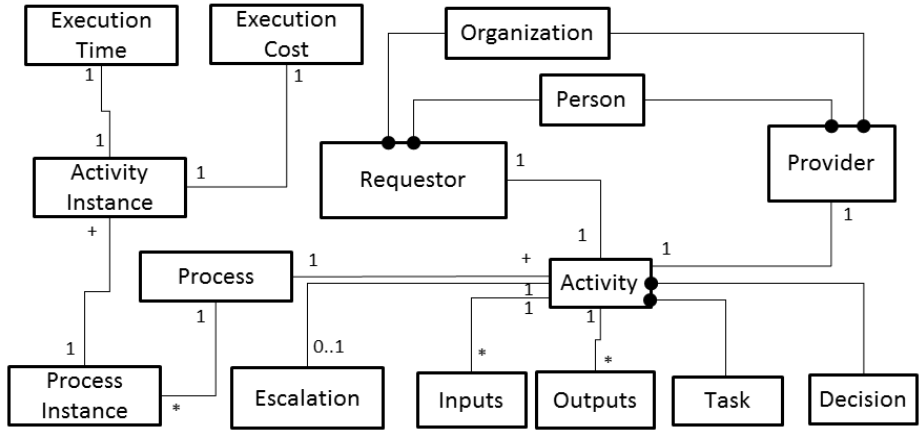}
    \caption{$\Aniti$ process meta-model for corruption.}
    \label{fig:metamodel}
\end{figure*}

We demonstrate the first step of the corruption-free vision in this section.
Given the corruption scenarios in the motivating example, we need a 
vocabulary that enables a computational analysis of corruption-prone process
models as well as process instances.  For this, we use 
the $\Aniti$ meta-model \cite{corruption-semcity13}  (see Figure~\ref{fig:metamodel}). 
Here, nodes represent concepts,
edges represent relationships and edges with circled ends denote generalization relationships.  

So, a process activity can
specialize as a decision or a task.  Each activity of a process is
requested by an entity and is provided by another.  The requester and
the owners could be individuals or organizations.  A task takes many
inputs, produces many outputs, and may have an escalation process
associated with it that can be triggered by the requester.  A process
also maybe instantiated many times, each instance capturing the
execution trace along with the time and cost of execution for each
activity. 

We now show how  the meta-model can be used to distinguish
corrupt process instances from uncorrupt ones. 
See \cite{corruption-semcity13} on how this meta-model is operationalized as a
first-order predicate language, used to define corruption patterns (i.e., monopoly, 
no SLA, discretion, lack of reviewability and lack of visibility) and then applied for  detecting them
based on the functional and non-functional characteristics of process models and instances.

\subsection{Detecting Corruption with  $\Aniti$}
\label{subsec:detectingcorruption}

Suppose we have a process $P$ $\in \mathbb{P}$, whose instance $i$
$\in \mathbb{I}^P$, when executed with service request (input data)
$o_{in}$ leads to a result $o_{out}$ (output data).  We define
\fsf{provenance$^*$ ($o_{in}$, $o_{out}$)} as the transitive closure
  of \fsf{provenance} to denote the input and output of an instance.
  Similarly, let \fsf{executionTimeOf$^*$ ($i$, $t$)} and
    \fsf{executionCostOf$^*$ ($i$, $c$)} be the time and costs of $i$
      derived from the time and cost of individual activities in $i$
      (e.g., as a sum of time and sum of cost).

Not all instances of $P$ witness corruption.  Suppose that $i_1 \in
\mathbb{I}^P$ has been uncorrupted while
$i_2 \in \mathbb{I}^P$ has been corrupted.  Given identical inputs,
the outputs of an uncorrupt process may differ from those of a corrupt
one.  For $i_1$ and $i_2$ to be comparable, they need to have
identical inputs.  In many cases, this may not be feasible since once
an input (e.g., service request for a driver license) has been
processed, the same input cannot be processed again for legal reasons.
In those cases, we assume that the inputs are approximately same,
i.e., similar, denoted by $\approx$.

Now considering \fsf{provenance$^*$ ($in_1$, $out_1$)} for instance
$i_1$ and \fsf{provenance$^*$ ($in_2$, $out_2$)} for instance $i_2$,
one can compare $out_1$ and $out_2$ with the aim to detect corruption.

\begin{itemize}
\item $out_1$ and $out_2$ are the same, but $out_2$ was obtained in
  much faster time than $out_1$. Hence, \fsf{executionTimeOf$^*$ ($i_1$,
  $t_1$)}, \fsf{execution TimeOf$^*$ ($i_2$, $t_2$)}, and $t_2 \ll t_1$.

\item $out_1$ and $out_2$ are the same, but $out_2$ was obtained by
  expending different resources than $out_1$. Hence,
  \fsf{executionCostOf$^*$ ($i_1$, $c_1$)},
    \fsf{executionCostOf$^*$ ($i_2$, $c_2$)}, and $c_1 \neq c_2$.  Here,
      the cost could be more or it could be less for the corrupt
      instance depending on who bears most of the cost.  If the
      requester bears the most of it, corrupt instances would observe
      less costs and the converse is true if the provider bears it.

\item $out_1$ and $out_2$ are different.  This means that even though
  the same inputs were given, the results were different. It indicates
  that (a) some decision (by an actor) was not consistently taken, or
  (b) some data was not considered by a processing step.
\end{itemize}

\subsection{Discussion}

In above, a process is deemed suspicious (possibly corrupt) if given two same or similar
 service requests (inputs), 
they lead to different outcomes and hence outputs (functionally or non-functionally).
It makes the assumption that corruption-free processes are unbiased, and in unbiased processes, 
similar inputs should lead to the similar outcomes in terms of their
results (functional outputs) and also the time and resources incurred to get those 
results (non-fuctional outputs). 

The benefit of this approach is that detecting 
variance in outputs is computationally easier than discovering an agent's suspicous gains.
However, this definition will not work if the the service request is unique because there
is not enough data to compare. 

To aid such a computable detection of corruption, governments should:
\begin{itemize}

\item Make raw and aggregate data about service requests publicly available. In doing so, they should
         anonymize sensistive or personally identifiable information which are irrelevant to the output
         of the service requests.
\item Setup baseline for performance and monitor performance data on service requests for aberrations. 
         The aberrations are candidates for corruption.
\item Use historical data on service performance to re-visit baselines periodically over time,
         alter processes steps or change actors.
\item define a transparent process to tackle unique service requests so that they can be later audited. 

\end{itemize}

%% file: challenges.tex

\section{Technical Challenges to Achieve Vision}
\label{sec:challenge}

We now look at some of the technical challenges which can help achieve the vision
of less-corruptible public processes.

\noindent {\bf 1. Formalizing Process Models for Corruption Analysis }

In order to represent and reason with public proceeses, we need a vocabulary of the key 
concepts in the corruption domain. 
Business process languages like BPMN\cite{bpmn} can serve as a starting point but needs 
to be extended. Here, there is work on patterns and anti-patterns for capturing
desirable service attributes\cite{patterns}. However, this has not been applied for 
detecting corruption in public services.

The importance of a formal model for corruption has long been felt but there is very little work.
One example we are aware of
is \cite{corruption-credit-formalmodel} which presents a model to connect informal and formal
credit markets. Another work is seen
in \cite{corruption-semcity13} where a meta-model for corruption  is formalized
using a first-order predicate language $\Aniti$. 
Compared with generic business process that capture activities, actors, inputs and outputs, execution
time and costs, the meta-model additionally captures escalation paths, 
data provenance, actor inter-relationship and task visibility. 


\noindent {\bf 2. Simulating processes}

Agent-based simulation is a flourishing discipline studying emergent behavior of large
scale systems. Public processes offer an attractive application area to study service request,
the corresponding service provider behavior, agent incentives and long-term process characteristics. 

Further, insights of agent societies will help understand how corruption spreads over time
and with more processes becoming corruptible.

\noindent {\bf 3. Agent Reasoning}

Given that actors may have varying degree of discretion in public processes, planning and reasoning
for the service requester and service provider become even more complex. As additional
challenge, one would
like to know if multiple actors could collude to abuse their positions to corrupt the process. 
This can happen among actors in service providers, among different service requesters or even
across them.

Agents cooperating with each other can also help in removing discretion and hence corruption. 
Some examples are transparent decision making techniques like voting protocols, auctions 
and negotiations. Another challenge will be to introduce  such techniques during process
design with minimal change(s).

\noindent {\bf 4. Agent Economics}

Since at the heart of corruption is an actor's desire to make unreasonable 
gain (which may or may not be observable), 
modeling incentives of individuals and the society at large will be challenging. Further, knowing 
the rewards of corrupt practice and the risk of punishment, incentives and disincentives of the actors 
at key decision points need exploration.

\noindent {\bf 5: Human and Agent Interaction}

Since public processes are essentially human centered, using agent techniques bring
challenges in human agent interaction into focus. Agents can analyze interactions of human actors to
learn relationships (like cartelling or conflict-of-interest). Agents can also interact with humans
to alert them on potential instaces of corruption or advise remedial actions.

\noindent {\bf 6: System Traceability}

A general challenge for improving process transparency is to provide  traceability information for
process instances executed. This becomes quite challenging if sub-parts of the process flow 
are outsourced to other organizations. Further, if an instance is found suspicious, we need an
ability to pin-point to likely causes using established service baselines.


%% file: solution.tex
\section{Discussion}
\label{sec:discussion}

In this section, we will illustrate that even initial steps to formalize public processes
can lead to better understanding of corruption. 
We  revisit the questions raised about corruption earlier and try
to answer them using $\Aniti$ \cite{corruption-semcity13}.

\begin{itemize}
\item Can a service for which the customer does not pay a fee
     (free service) be corruptible? {\bf Answer}: The service can still be
		corrupt because, given similar input, their execution times may be different
		or even their service outcome may be different.
\item Can a service be corrupt if the agent is not a human? {\bf Answer}: The service 
    can be corrupt since nature of the agent has no relevance to difference in outcome
		of two similar service requests. 
\item Can corruption happen if delivering a service  requires no contention of resources?    
    {\bf Answer}: The service can be corrupt since availability of resources is
		irrelevant to the cost, time or results of two similar service requests.
\item Can a service be corrupt if it is one-of-a-kind and hence
         only requested once?  {\bf Answer}: The $\Aniti$ model only detects corruption by comparing
         outcomes of similar service requests. For a unique request, it cannot detect corruption.
\item If a service requester's and provider's familiarity to each other are known, does it promote 
         corruption? 
        {\bf Answer}: The $\Aniti$ model can represent actor inter-relationships and thus can
     detect conflict-of-interest. However, it will only be able to
     detect corruption if discrepancy is seen among outcome for similar service requests. 
\item If a service request was found to be handled in a corrupt manner, 
        can the culpability of agents involved be established legally?
        {\bf Answer}: If provenance information on service provider's side is maintained, all responsible 
       actors can be determined. The legal ramifications are unknown.
\item Does allowing brokers to mediate service requests on behalf of service requesters
         promote corruption? {\bf Answer}: The impact of brokers in corruption can be detected
         by comparing outcomes of similar service requests with and without brokers. To understand 
         the role of brokers, modeling of their relationships with  actors in service delivery is needed 
         which the model allows.
\end{itemize}

The questions were not exhaustive but an initial formalization of the corruption domain helps
answer some of them. 

%% file: conclusion.tex

\section{Conclusion}
\label{sec:conclusion}

In this paper, we painted a vision of how ICT in general and agent technologies in particular
can help tackle corruption in public services and identifed challenges to achieve the vision.
We illustrated the vision using the common public process of national registeration of citizens
seen around the world. Although some work is seen in detecting corruptions, a lot more
needs to be done to formalize, detect, address  and ultimately prevent corruption.
